\documentclass[conference,a4paper]{IEEEtran}

\usepackage{color}

\usepackage{balance} 

\usepackage{cite}

\usepackage[hidelinks]{hyperref}

  \usepackage{graphicx}
  
  \DeclareGraphicsExtensions{.eps}

\begin{document}


\title{Economic Value of Inertia\\in Low-Carbon Power Systems}

\author{\IEEEauthorblockN{Luis Badesa, Fei Teng and Goran Strbac}
\IEEEauthorblockA{Department of Electrical and Electronic Engineering\\
Imperial College London\\
London SW7 2AZ, United Kingdom\\
Email: \{luis.badesa, f.teng, g.strbac\}@imperial.ac.uk}}

\maketitle

\begin{abstract}

Most renewable energy sources (RES) do not provide any inertial response. Their integration in a power grid implies a highly reduced level of system inertia, which leads to a deteriorated frequency performance. Then, the requirement for frequency response is significantly increased in order to maintain frequency security. Alternatively, enhanced provision of inertia from auxiliary sources may alleviate this problem. However, the benefits of inertia provision are not yet fully understood.

In this paper, an inertia-dependent Stochastic Unit Commitment (SUC) tool is applied to quantify the economic value of inertia. The results demonstrate that enhanced provision of inertia would lead to significant economic savings, although these savings vary under different system conditions. These results should be brought to the attention of both market operators and investors, in order to inform the design of an ancillary-services market for inertia and the investment in auxiliary provision of inertia.

\end{abstract}


\section*{Nomenclature}
\addcontentsline{toc}{section}{Nomenclature}

\subsection*{Indices}
\begin{IEEEdescription}[\IEEEusemathlabelsep\IEEEsetlabelwidth{$RoCoF_{max}$}]
\item[$g$] Thermal generator.
\item[$n$] Node number.
\end{IEEEdescription}

\subsection*{Sets}
\begin{IEEEdescription}[\IEEEusemathlabelsep\IEEEsetlabelwidth{$RoCoF_{max}$}]
\item[$\mathcal{G}$] Set of thermal generators.
\item[$\mathcal{N}$] Set of nodes in the scenario tree.
\end{IEEEdescription}

\subsection*{Constants}
\begin{IEEEdescription}[\IEEEusemathlabelsep\IEEEsetlabelwidth{$RoCoF_{max}$}]
\item[$\Delta\tau(n)$] Time-step corresponding to node $n$ (h).
\item[$\Delta f_{max}$] Maximum admissible frequency deviation from nominal value (Hz).
\item[$\pi(n)$] Probability of reaching node $n$.
\item[$c^{m}_g$] Marginal cost of thermal unit $g$ (\pounds/MWh).
\item[$c^{nl}_g$] No-load cost of thermal unit $g$ (\pounds/h).
\item[$c^{st}_g$] Startup cost of thermal unit $g$ (\pounds).
\item[$D$] Load damping rate (1/Hz).
\item[$f_0$] Nominal frequency level (Hz).
\item[$H_g$] Inertia constant of thermal unit $g$ (s).
\item[$H_L$] Inertia constant of tripped generator (s).
\item[$H_L^{max}$] Inertia constant of largest generation unit (s).
\item[$P_{g}^{max}$] Maximum generation of thermal unit $g$ (MW).
\item[$P_L$] Generation loss (MW).
\item[$P_L^{max}$] Generation loss of largest unit (MW).
\item[$RoCoF_{max}$] Maximum rate of change of frequency admissible (Hz/s).
\item[$T_d$] Delivery time of Primary Frequency Response (s).
\end{IEEEdescription}

\subsection*{Semi-Constants (fixed with respect to linear program but variable between time-steps)}
\begin{IEEEdescription}[\IEEEusemathlabelsep\IEEEsetlabelwidth{$RoCoF_{max}$}]
\item[$P^D(n)$] Total demand at node $n$ (MW).
\end{IEEEdescription}

\subsection*{Decision Variables}
\begin{IEEEdescription}[\IEEEusemathlabelsep\IEEEsetlabelwidth{$RoCoF_{max}$}]
\item[$P_g(n)$] Power output of thermal unit $g$ at node $n$ (MW).
\item[$R_g(n)$] Primary Frequency Response provision from thermal unit $g$ at node $n$ (MW).
\end{IEEEdescription}

\subsection*{Linear Expressions (linear combinations of decision variables)}
\begin{IEEEdescription}[\IEEEusemathlabelsep\IEEEsetlabelwidth{$RoCoF_{max}$}]
\item[$C_g(n)$] Operating cost of thermal unit $g$ at node $n$ (\pounds).
\item[$H(n)$] System inertia after generation loss at node $n$ (MW$\cdot s^2$).
\item[$N_g^{sg}(n)$] Number of thermal unit $g$ that start generating at node $n$.
\item[$N_g^{up}(n)$] Number of thermal unit $g$ that are online at node $n$.
\item[$R(n)$] Total frequency-response provision at node $n$ (MW).
\end{IEEEdescription}

%
\IEEEpeerreviewmaketitle

\section{Introduction}


Decarbonisation of power systems introduces significant challenges, being one of the most significant the deterioration of frequency performance. This deterioration is mainly caused by the inertia reduction that comes associated with integration of most RES. The reduced system inertia increases the required amount of frequency response or the risk of violating frequency security.

Assuring frequency security means that the system's frequency must be maintained within a relatively narrow range, as generating units could be damaged if these frequency limits were not respected. Frequency deviations from steady state occur constantly in a real power grid, and are caused by an imbalance between generation and demand of energy. Most of these imbalances are minuscule, and are automatically corrected by the action of control devices distributed around the grid. However, if there is a generation outage in the system, frequency will drop rapidly and significantly, risking security. The rate at which the frequency drops is determined by the inertia of all the generators and rotating loads connected to the system. 

In order to alleviate the challenges caused by this reduced level of inertia, some potential solutions are currently being analysed by the research community. Enhanced Frequency Response (EFR) and Demand Side Response (DSR) are the most promising ones. As an alternative, cost-effective provision of extra inertia from auxiliary sources could also support frequency control. The extra inertia could be provided by thermal generators, rotating loads, or even wind generators. Thermal generators and rotating loads could be manufactured with a higher inertia constant, while wind generators could potentially provide ``synthetic inertia" through a supplementary control loop in the wind turbine controller. 
However, the current market mechanisms do not reward the provision of inertia and therefore there is lack of incentive for investors to develop as alternative providers of inertia. Reference \cite{ElaI} pointed out the potential benefits of implementing a market for inertia. This paper hence focuses on understanding the value of inertia provision and informs the design of market mechanisms for inertia.

In fact, very few studies have analysed the economic value of inertia. The only paper that has attempted to address this issue is \cite{OverbyeInertiaValue}, in which a security-constrained Unit Commitment (UC) framework is used to quantify the value of inertia in a system with wind generation. However, some assumptions made by this work might lead to imprecise conclusions. A load damping factor of zero is considered, which could lead to over-estimating the value of inertia. A deterministic UC model is used to quantify the value of inertia, which may give inaccurate results when considering wind generation due to the stochastic nature of wind. Furthermore, although the paper uses a security-constrained UC model, only the maximum frequency deviation is assured to be maintained under acceptable limits. The maximum Rate of Change of Frequency (RoCoF) and quasi-steady-state requirements are not considered. The RoCoF requirement is particularly important when assessing the value of inertia because RoCoF is limited only by scheduling inertia, as will be discussed in Section III of this paper. Ignoring the RoCoF requirement may lead to an underestimation of the value of inertia.

Therefore, there is a clear need to accurately assess the value of inertia in power systems. In the present paper, a Stochastic Unit Commitment (SUC) tool is used to quantify the economic value of inertia in a power system with significant wind penetration. This study is of interest to both energy market operators and market participants. Its aim is to inform market operators of the necessity for an ancillary market for inertia, by quantifying how much they should incentivise inertia provision depending on the system's characteristics and conditions. As well, manufacturers of generating plants will be able to assess the revenue that making generators with higher inertia would bring, given that the appropriate market framework is put in place. 

The SUC model used to conduct the study is described in Section II. Section III gives the dynamic model of frequency evolution considered in this paper. The results of the study are discussed in Section IV, while Section V gives the conclusion.

\section{Stochastic Scheduling Model}

In this paper, the economic value of inertia is defined as the savings in operational cost of the system when its level of inertia is increased. In order to calculate the cost associated with the operation of a power system, the present study makes use of the multi-stage stochastic scheduling model developed in \cite{AlexEfficient} and expanded in \cite{FeiStochastic}. This tool optimally schedules energy production and delivery of several ancillary services, while considering the uncertainty introduced by wind generation. 

\begin{figure}[!t]
\vspace{3mm}
\centering
\includegraphics[width=3.3in]{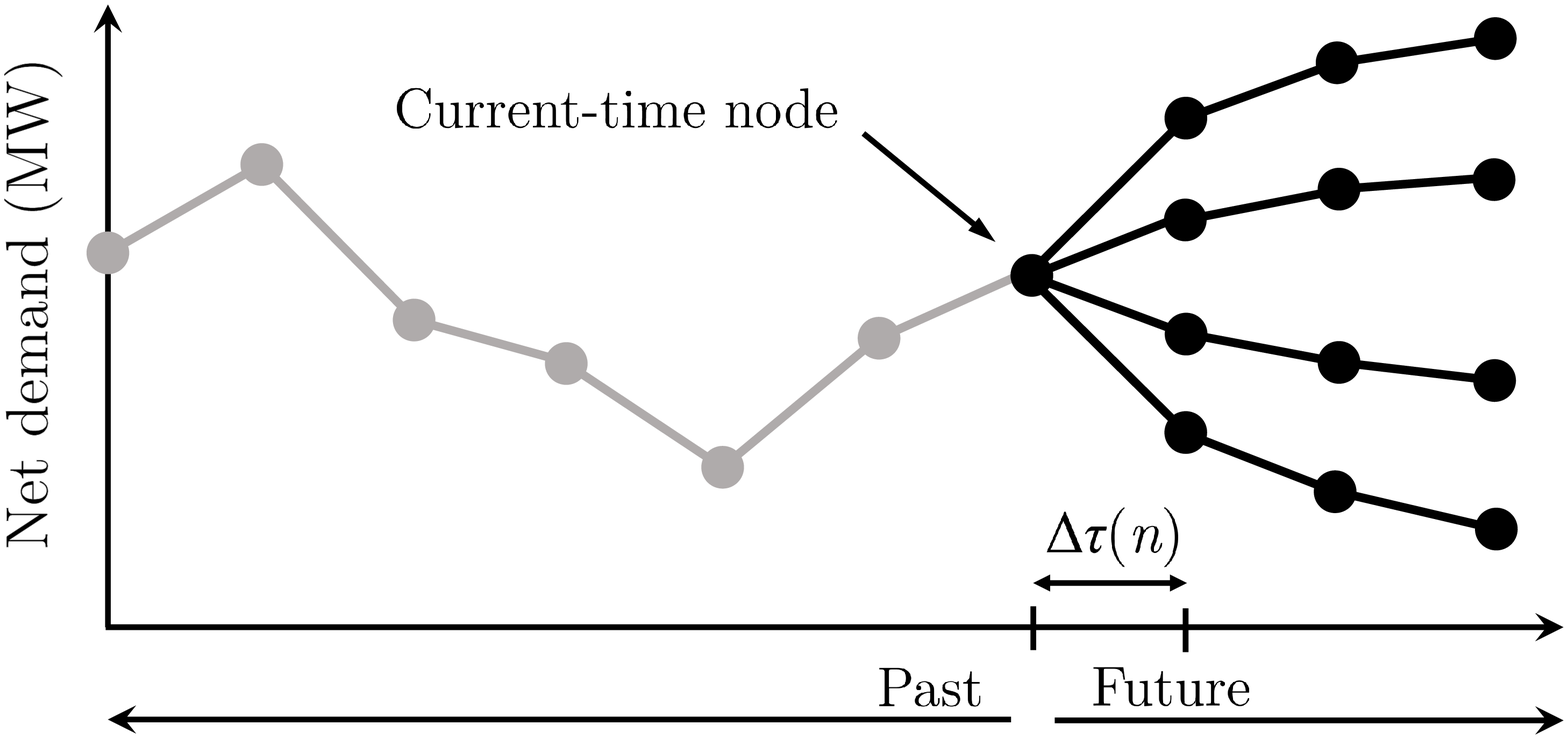}
\vspace{3mm}
\caption{Example of a scenario tree used in the Stochastic Unit Commitment}
\label{ScenarioTree}
\end{figure}

The optimisation problem in the SUC is solved over a multi-stage scenario tree as the one shown in Fig. \ref{ScenarioTree}. The scenario tree is built using the quantile-based, scenario-selection method described in \cite{AlexEfficient}. Each scenario corresponds to a user-defined quantile of the distribution of net demand. Net demand is defined as the difference between load and forecast wind generation for future nodes, while it is defined as the difference between load and realised wind generation for past and current nodes. For simplicity, scenario trees are constructed in this paper with branching only at the current-time node. This approach provides similar results to more intricate structures while greatly reducing computational time, as demonstrated in \cite{AlexEfficient}.

Once the scenario tree is defined, a rolling-planning approach is used. First, a full SUC calculation is completed with a 24-h horizon in an hourly time-step, discarding all decisions after the here-and-now ones. In the next time-step, realisations of the stochastic variable, wind generation, become available, as well as new wind forecasts. An updated scenario tree is then built, covering again a 24-hour time horizon, and the SUC calculation is run again. Inter-temporal constraints are maintained in this new scenario tree, while decision variables are adjusted after solving the new optimisation problem. This methodology, consisting on building new scenario trees for each time-step, is repeated for the whole duration of the simulation, hence the term rolling-planning.

The goal of the SUC optimisation problem, formulated as a mixed-integer linear program (MILP), is to minimise the expected operation cost over all nodes in the scenario tree. The objective function of the SUC is:
\begin{equation} 
\sum_{n \in \mathcal{N}}\pi(n)\sum_{g \in \mathcal{G}}C_g(n)
\end{equation}
The sum of the operating cost of all thermal units at each node, which is the total operating cost at that node, is weighted by the probability of reaching that node, $\pi(n)$. The calculation of this probability associated to node $n$ is derived from the user-defined quantiles, using the procedure defined in \cite{AlexEfficient}. The operating cost for generator $g$ is given by:
\begin{equation} 
C_g(n)=c_g^{st}N_g^{sg}(n)+\Delta\tau(n)\left[c_g^{nl}N_g^{up}(n)+c_g^{m}P_g(n)\right]
\end{equation}

This objective function is subject to several constraints, in order to correctly model the behaviour of a power system and assure system security as well. Two of the security-related constraints will be discussed in the next section, while a comprehensive mathematical description of the SUC model can be found in the appendix of \cite{FeiStochastic}. The SUC model considers a single-bus representation of the system, in which transmission constraints are not modelled.

\vspace{3mm}
\section{Dynamic Model of Frequency Evolution} \label{SectionDynamicFrequency}
The uniform frequency model is adopted in this study, which assumes that all the generators move coherently as a single lumped mass and load damping is modelled as a single constant. Therefore, the dynamics of frequency deviation after a generation outage in the power system are represented by a single differential equation \cite{KundurBook}:
\begin{equation} \label{KundurEq}
2H\frac{d \Delta f(t)}{d t}+D\cdot P^D\cdot\Delta f(t)=\sum_{g \in \mathcal{G}}\Delta P_g(t)-P_L
\end{equation}
where $\Delta P_g$ [MW] is the additional power provided by thermal unit $g$ following the generation loss $P_L$. Term $\Delta P_g(t)$, also referred to as delivery of Primary Frequency Response (PFR), is modelled as:
\begin{equation} \label{PFReq}
\Delta P_g(t)=\left\{ 
\begin{array}{ll}
\frac{R_g}{T_d}\cdot t \quad & \mbox{if $t<T_d$} \\
R_g \quad & \mbox{if $t\geq T_d$}
\end{array}
\right.
\end{equation}
In (\ref{KundurEq}), (\ref{PFReq}) and following equations, index $n$ has been dropped from the decision variables, linear expressions and semi-constants for brevity.

The system's level of inertia after the generation loss, $H$, is given by:
\begin{equation}
H=\frac{\sum_{g \in \mathcal{G}}H_g\cdot P_g^{max}\cdot N_g^{up}-P_L\cdot H_L}{f_0}
\end{equation}

\vspace{2mm}
\subsection{Dynamic Frequency Constraints}
In order to assure a secure operation of the system from a frequency-performance point of view, system operators must make sure that three requirements are met: first, RoCoF must never exceed a certain limit; second, the frequency nadir (minimum value of frequency) must always be above a certain value; and third, the frequency deviation in quasi-steady-state must be below a predefined threshold. Only the RoCoF and nadir requirements are discussed in this paper, as quasi-steady-state deviation is not affected by inertia. For a description of the quasi-steady-state requirement, refer to \cite{FeiStochastic}.

When defining the dynamic frequency constraints, this paper considers only the loss of the largest generation unit, as this is the most severe case and thus will provide the most severe constraints. Therefore, $P_L$ and $H_L$ are substituted from now on by $P_L^{max}$ and $H_L^{max}$, respectively.

The RoCoF requirement is given by:
\begin{equation} \label{ROCOFeq}
H\geq\left|\frac{P_L^{max}}{2\cdot RoCoF_{max}}\right|
\end{equation}
which can be obtained from (\ref{KundurEq}) by noticing that RoCoF takes its maximum value in the instant after the generation loss occurs, considering that PFR provision is negligible at that instant. 

The nadir requirement is given by:
\begin{equation}
H\cdot R\geq k^{*} \label{NadirEq}
\end{equation}
Where $k^{*}$ is a function of several parameters:
\begin{equation} \label{kEq}
k^{*}=f(\Delta f_{max},\,P^D,\,D,\,T_d,\,P_L^{max})
\end{equation}
It is important to point out that $k^{*}$ is a decreasing function of $P^{D}$, as load damping supports managing the post-fault frequency condition. The rest of the arguments in (\ref{kEq}) would be constant for a particular system. For a detailed deduction of this constraint, refer to \cite{FeiStochastic}. As this is a bilinear constraint, it must be linearised before being implemented in an MILP formulation. The linearisation procedure used in this study is also described in \cite{FeiStochastic}.

An increased system inertia would induce a decrease in operational costs, as the algorithm would not need to schedule as much inertia to comply with (\ref{ROCOFeq}) and (\ref{NadirEq}). Further discussion on this reasoning is included in the following section, making use of examples to better illustrate our point. 

\vspace{3mm}
\section{Assessment of the Value of Inertia}
This section presents the results and analysis of the case studies. The characteristics of the generation mix used for this work are included in Table \ref{TableThermal}. The load damping factor, $D$, was set to 0.5\%. A scenario tree branching only in the current-time node was used. Net demand quantiles of 0.005, 0.1, 0.3, 0.5, 0.7, 0.9 and 0.995 were considered for that scenario tree. A penalty for emissions of 150 \pounds /ton$\mbox{CO}_{2}$ was also considered.

\begin{table}[!t]
\renewcommand{\arraystretch}{1.2}
\caption{Characteristics of Thermal Plants}
\label{TableThermal}
\centering
\begin{tabular}{l| l l l}
\hline
& Nuclear & CCGT & OCGT\\
\hline
Number of Units & 6 & 110 & 30\\
Rated Power (MW) & 1800 & 500 & 200\\
Min Stable Gen (MW) & 1800 & 200 & 50\\
No-Load Cost (\pounds/h) & 0 & 7809 & 8000\\
Marginal Cost (\pounds/MWh) & 10 & 51 & 110\\
Startup Cost (\pounds) & n/a & 9000 & 0\\
Startup Time (h) & n/a & 4 & 0\\
Min Up Time (h) & n/a & 4 & 0\\
Min Down Time (h) & n/a & 1 & 0\\
Inertia Constant (s) & 5 & 5 & 5\\
Max Response (MW) & 0 & 50 & 20\\
Response Slope & 0 & 0.5 & 0.5\\
Emissions (kg$\mbox{CO}_{2}$/MWh) & 0 & 368 & 833\\
\hline
\end{tabular}
\end{table}

\subsection{Annual Value}
\begin{figure}[!t]
\centering
\includegraphics[width=2.7in]{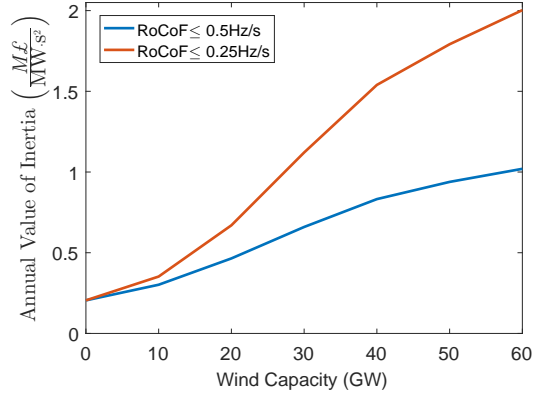}
\caption{Annual value of inertia for different levels of wind capacity.}
\label{AnnualWind}
\end{figure}

The annual value of inertia under different penetration levels of wind generation is presented here. Several simulations considering a one-year-long operation of the system were conducted, with and without extra freely-available inertia. The value of inertia was calculated as the reduction in operational cost for the cases with freely-available inertia. The results of the study are presented in Fig. \ref{AnnualWind}. 

Fig. \ref{AnnualWind} shows two different cases: simulations run with a $RoCoF_{max}$ of 0.5Hz/s and simulations run with a $RoCoF_{max}$ of 0.25Hz/s. The RoCoF requirement of 0.5Hz/s is a relaxed requirement from the current $RoCoF_{max}$ of 0.25Hz/s imposed by National Grid in the Great Britain system \cite{StandardNationalGrid}. Relaxing this requirement was proposed by \cite{ReportROCOF}, and \cite{FeiStochastic} demonstrated that a relaxed $RoCoF_{max}$ would significantly reduce the system's operational cost. Therefore, both cases have been considered in the present paper. Fig. \ref{AnnualWind} shows that relaxing the RoCoF requirement would reduce the value of inertia. Nevertheless, inertia would still have a significant value in a system with a $RoCoF_{max}$ of 0.5Hz/s.

As can be seen in Fig. \ref{AnnualWind}, the value of inertia monotonically increases with installed capacity of wind generation. In the case of low installed capacity of wind, the value of inertia is relatively low. This is the case in the present British system, which is still dominated by conventional plants, hence there is no need to establish a market for inertia. 
However, in the case of 60GW of wind capacity, one unit of inertia (MW$\cdot \mbox{s}^2$) would reduce the annual system operation cost by more than 1M\pounds, for a $RoCoF_{max}$ of 0.5Hz/s. 

In order to analyse the relative savings due to addition of inertia, let's consider a particular example. For the case of 30GW of wind capacity and a $RoCoF_{max}$ of 0.5Hz/s, the annual operational cost of the system is of 12.22B\pounds. The value of inertia for that same case is of 659.5k\pounds, as can be observed in Fig. \ref{AnnualWind}. This might seem as an insignificant reduction in cost. Nevertheless, this is the value of just one extra unit of inertia, so the savings would be much higher if the extra inertia in the system was further increased.

It has been demonstrated that provision of extra inertia leads to a reduction in operational costs. This should be recognised and rewarded, so that the investors would be incentivised to invest in alternative sources for the provision of inertia. Under a market that properly rewards the provision of inertia, the results in Fig. \ref{AnnualWind} can be used to inform the investment. 

\subsection{Instantaneous Value}

\begin{figure}[!t]
\centering
\includegraphics[width=3.2in]{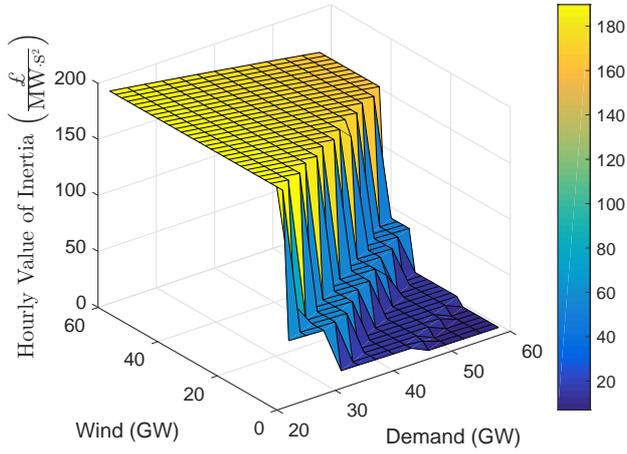}
\caption{Instantaneous value of inertia for a RoCoF requirement of 0.5Hz/s.}
\label{Instantaneous1}
\end{figure}

\begin{figure}[!t]
\centering
\includegraphics[width=3.2in]{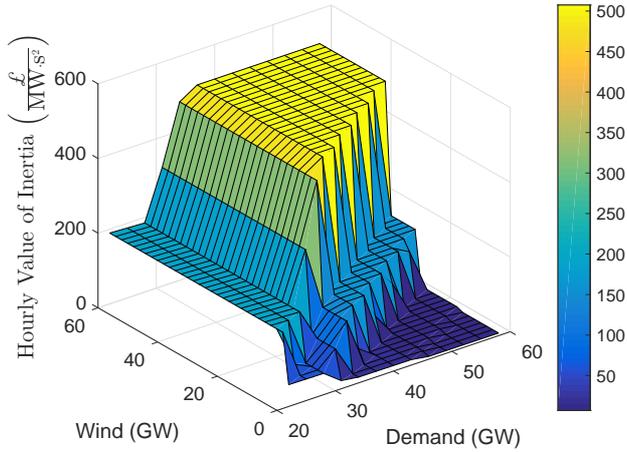}
\caption{Instantaneous value of inertia for a RoCoF requirement of 0.25Hz/s.}
\label{Instantaneous2}
\end{figure}

In addition to annual benefits from extra inertia provision, it is also important to recognise that the value of inertia may vary significantly from hour to hour, depending on the system's conditions. Therefore, we further analyse the hourly economic savings that an additional unit of inertia in the system would deliver, for a particular demand-wind condition. The results are presented in Figs. \ref{Instantaneous1} and \ref{Instantaneous2} for a $RoCoF_{max}$ of 0.5Hz/s and 0.25Hz/s, respectively.

As shown in Fig. \ref{Instantaneous1}, the value of inertia is directly linked with the level of net demand (difference between demand and wind generation). Two distinctive areas can be observed: an area with a low value for inertia, but with an increasing trend, located in the high-demand, low-wind region of the graph; and a relatively flat area with a high value for inertia, located in the high-wind region. The difference between these two areas can be explained if we consider the effect of wind curtailment. When net demand is low, the frequency-response requirement is high due to the low level of system inertia (note that nadir is the binding constraint, as explained further on). Then, wind generation has to be curtailed in order to maintain enough conventional plants online, as they need to provide sufficient inertia and frequency response to assure frequency security. In those cases, adding extra inertia allows for less wind to be curtailed, therefore highly reducing the operational costs as wind generation has zero marginal cost and zero $\mbox{CO}_{2}$ emissions. This situation corresponds to the high-wind region of the graph. One can also notice that the value of inertia in this high-wind region slowly decreases for increasing demand. This result can be explained by the fact that the nadir requirement  (\ref{NadirEq}) is the binding constraint, as the increased load damping effect described in (8) lowers the requirement for $H$ and $R$ for higher levels of demand.


The other distinctive area in Fig. \ref{Instantaneous1} is related to cases with high net demand, in which no wind generation would be curtailed. In those cases, corresponding to the high-demand, low-wind region, the extra inertia still has some value, but one order of magnitude lower than in the wind-curtailment cases. For these high-net-demand cases, extra inertia reduces the requirement for provision of frequency response $R$,  allowing a smaller number of thermal units to run in the partially-loaded mode. This effect slightly reduces the operational cost of the system. 

This same study was repeated for a $RoCoF_{max}$ of 0.25Hz/s and the results are presented in Fig. \ref{Instantaneous2}. The two distinctive areas in Fig. \ref{Instantaneous1} can also be seen in Fig. \ref{Instantaneous2}: a flat area with a high value for inertia, located in the high-demand, high-wind region; and an area with a low value for inertia, located in the high-demand, low-wind region of the graph. The value of inertia is much higher in both these areas, when compared to those same regions in Fig. \ref{Instantaneous1}. This higher value is due to RoCoF being the binding constraint for a level of demand higher than 30GW in Fig. \ref{Instantaneous2}. As load damping helps lower the nadir requirement for high levels of demand and the RoCoF requirement has not been relaxed in this case (remember that $RoCoF_{max}=$ 0.25Hz/s), RoCoF becomes the binding constraint for demand levels higher than 30GW. This is graphically demonstrated in Fig. \ref{Instantaneous2}: the value of inertia in the yellow area does not decrease for increasing demand, meaning that nadir is not the binding constraint (no load damping effect is present). 

Note that the value of inertia in Fig. \ref{Instantaneous2} is highest in the yellow area of the graph. As RoCoF is the binding constraint there and net demand is very low, a high number of conventional plants are scheduled only to provide inertia. Extra inertia provision for those wind-demand conditions would directly reduce the number of online plants and therefore significantly reduce the curtailment of wind, which explains the very high value of inertia in that region. Notice however that for levels of demand lower than 30GW, inertia has the same value as in the relaxed RoCoF case (shown in Fig. \ref{Instantaneous1}). That is because nadir is the binding constraint in that region, as the effect of load damping is not very significant due to the low level of demand. In summary, inertia is more valuable when RoCoF is the binding constraint, as RoCoF directly depends on the system's level of inertia $H$, while nadir depends on the product $H\cdot R$.

The main conclusion that can be made from these results is that the instantaneous value of the system must be carefully studied for each power system before designing an adequate inertia market, as there are many nuances. However, a key result that holds true for any system, regardless of which constraint is binding, is that inertia is most valuable when it allows for wind curtailment to be reduced.

The study presented in this subsection serves as a starting point for informing the design of an hourly inertia market, as it clearly demonstrates the time-varying value of inertia provision, depending on the different system's conditions.

\begin{figure}[!t]
\centering
\includegraphics[width=2.5in]{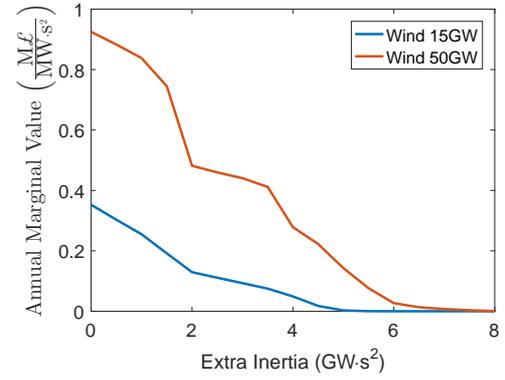}
\caption{Annual marginal value of inertia, for a RoCoF requirement of 0.5Hz/s.}
\label{MarginalAnnual_05}
\end{figure}

\begin{figure}[!t]
\centering
\includegraphics[width=2.5in]{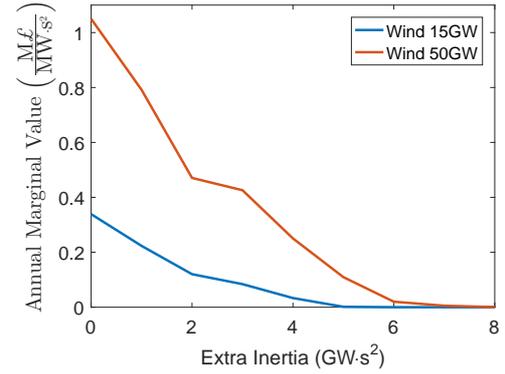}
\caption{Annual marginal value of inertia, for a RoCoF requirement of 0.25Hz/s}
\label{MarginalAnnual_025}
\end{figure}

\subsection{Marginal Value}

The previous two studies have demonstrated the value of adding an extra unit of inertia in the system. This subsection will focus on analysing how much extra inertia the system would need. Fig. \ref{MarginalAnnual_05} shows the results for a one-year-long simulation, considering two levels on wind penetration: 15GW, the current wind capacity of the Great Britain network \cite{WindCapacityUK}; and 50GW, a possible future scenario with high wind penetration. A $RoCoF_{max}$ of 0.5Hz/s has been considered for that study. The results of the same study, but considering a $RoCoF_{max}$ of 0.25Hz/s, are shown in Fig. \ref{MarginalAnnual_025}.

\begin{figure}[!t]
\centering
\includegraphics[width=2.5in]{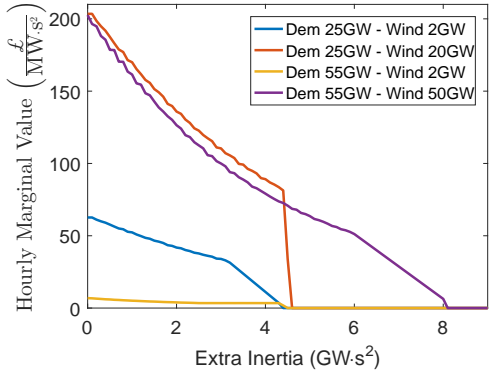}
\caption{Instantaneous marginal value of inertia, for a RoCoF requirement of 0.5Hz/s.}
\label{MarginalHourly_05}
\end{figure}

\begin{figure}[!t]
\centering
\includegraphics[width=2.5in]{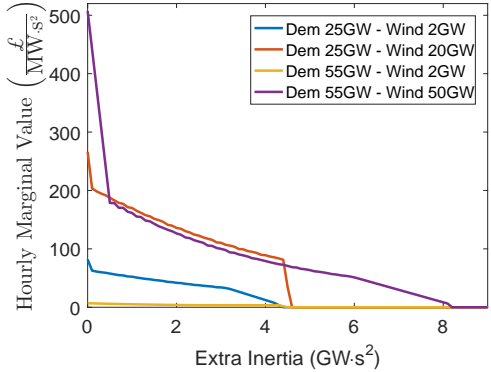}
\caption{Instantaneous marginal value of inertia, for a RoCoF requirement of 0.25Hz/s}
\label{MarginalHourly_025}
\end{figure}

The marginal value presented in these figures has been calculated as the difference between economic savings from adding a certain amount of extra inertia to the system, and savings from adding the immediately lower amount of extra inertia. As can be seen in both Fig. \ref{MarginalAnnual_05} and \ref{MarginalAnnual_025}, the marginal value declines along with increased provision of inertia, and saturates when more than 5$\mbox{GW}\mbox{s}^2$ are added to the 15GW-wind-capacity case, or 7$\mbox{GW}\mbox{s}^2$ to the 50GW-wind-capacity case. 

\balance 

The instantaneous marginal value of inertia was also studied, for several demand-wind conditions. The results are shown in Figs. \ref{MarginalHourly_05} and \ref{MarginalHourly_025}. As well as for the annual study, the marginal value declines along with an increased provision of inertia, while the saturation points vary depending on the different system conditions. The curves shown in these figures are not smooth, which is due to the linearisation of the bilinear nadir constraint, as discussed in Section \ref{SectionDynamicFrequency}. However, the general trend can be extracted from these results, and one conclusion can be made: there is no value in adding inertia above a certain saturation threshold. The saturation threshold is determined by the level of wind generation.

The results presented in this subsection need to be taken into consideration when developing market mechanism for inertia, because the value of inertia would start decreasing as investors decide to provide higher amounts of extra inertia.

\vspace{3mm}
\section{Conclusion}
This paper presents a quantification of the value of inertia in a low-carbon electric grid. The value of inertia was analysed from three different points of view: the annual value for different scenarios of wind penetration, the instantaneous value for different demand-wind conditions, and the marginal value. 

The results show that significant savings in the operational cost of the system could be achieved if provision of extra inertia was incentivised, particularly for cases of high wind generation. These results support the idea proposed by \cite{ElaI} that an inertia market should be designed. In particular, an hourly inertia market needs to be developed as the results clearly demonstrate the time-varying value of inertia provision, depending on the different system conditions. Under such a market framework, the results presented in the present paper would serve to inform both market operators and investors on provision of inertial response.

\vspace{3mm}

\bibliographystyle{IEEEtran} 
\bibliography{References_Inertia}

\end{document}